\documentstyle[preprint,aps]{revtex}
\begin{document}
 
\title{From $K^+$ in heavy-ion collisions to $K^-$ in kaonic atoms}
\bigskip
\author{G. E. Brown}
\address{Department of Physics, State University of New York,
Stony Brook, NY 11794, USA}
\author{C. M. Ko and G. Q. Li}
\address{Cyclotron Institute and Physics Department,\\
Texas A\&M University, College Station, Texas 77843, USA}
 
\maketitle
 
\begin{abstract}
Kaon production and kaon flow in heavy-ion collisions at SIS energies
are analyzed using the relativistic transport model.  The experimental 
data are found to be consistent with the scenario that kaon scalar 
(before being cut down by the range term) and vector mean field 
potentials are about 1/3 of that of nucleons in the Walecka model.
This also explains the strong attraction found for $K^-$ in kaonic
atoms, and gives information on the possibility of $K^-$ condensation 
in dense matter.
\end{abstract}
 
 
\section{introduction}

The study of kaon properties in dense matter is one of the most
interesting topics in nuclear physics. In ultrarelativistic heavy-ion 
collisions, enhanced production of kaon or other strange particles 
has been proposed as a possible signature for quark-gluon plasma 
formation in the collisions as the production rate in the quark-gluon 
plasma has been thought to be order of magnitude higher than in hadronic 
matter \cite{REF82}.  But the strangeness production rate in hadronic 
matter is usually estimated using free hadron properties. If the 
properties of the kaon or other strange particles are modified, then 
their production rate could be different.  Indeed, it has been pointed 
out in Refs. \cite{KO91,BROWN92} that if the kaon mass is reduced
in medium, then its production rate will be significantly enhanced.
Also, it has been suggested that the kaon yield from heavy-ion collisions 
at subthreshold energies is sensitive to the nuclear equation of state 
at high densities \cite{KO85} as kaons can not be absorbed due to 
strangeness conservation. However, to extract information on the 
nuclear equation of state from the experimental data requires transport 
model calculations in which the kaon in-medium properties are properly 
treated.  Furthermore, if $K^-$ feels a strong attractive potential as 
suggested by chiral perturbation theory and supported by the kaonic 
atom data, then it is possible that kaon condensation can occur in 
neutron stars \cite{BROWN88,PRA95,FUJ96}, leading to the possible 
formation of mini-black holes in galaxies \cite{BETHE94}.

Recently, Brown and Rho \cite{BROWN96A} have shown via an effective 
chiral Lagrangian that, with dropping pion decay constant in medium, 
the attractive scalar potential (before being cut down by the
energy-dependent range term) and the repulsive vector potential 
acting on a kaon are just 1/3 of nucleon mean-field potentials in 
Walecka model.  In this paper, we shall reexamine kaon production 
and flow in heavy-ion collisions at SIS energies using the relativistic 
transport model, which has been previously generalized to include the 
kaon mean-field potential, to see if the experimental data can
still be consistently explained by the potentials suggested in 
Ref. \cite{BROWN96A}.

The paper is organized as follows: In Section II, we shall review 
the kaon in-medium properties following the ideas in Ref. \cite{BROWN96A}. 
Then, kaon production and flow in heavy-ion collisions at SIS energies 
are studied in Section III. In Section IV, the antikaon potential is 
discussed in relation to the kaonic atom data \cite{FGB93}. Finally, 
conclusions are given in Section V.

\section{kaon in nuclear medium}

Since the kaon-nucleon $(KN)$ interaction is relatively weak when 
compared to other hadron-nucleon interactions, the impulse approximation 
should be reasonable for determining the kaon potential in nuclear matter 
at low densities. In this approximation, the kaon potential is directly 
related to the $KN$ scattering length in free space, i.e.,
\begin{eqnarray}\label{impulse}
U_{K^+} = - {2\pi \over m_K} (1+{m_K\over m_N}) {\bar a}_{KN} \rho ,
\end{eqnarray}
where $m_K$ and $m_N$ are the kaon and nucleon masses, respectively,
and ${\bar a}_{KN} \approx -0.255 $ fm is the isospin-averaged
KN scattering length in free space \cite{BARN94}. At normal nuclear
matter density $\rho _0 = 0.16 $ fm$^{-3}$, the kaon potential
is repulsive and has a magnitude of about 30 MeV.
 
Kaon properties in dense matter have been extensively studied in
chiral perturbation theory \cite{KAPLAN86,POLI91,BROWN94,LEE95}.
The most straightforward interaction on a kaon in dense matter
is the vector interaction, the so-called Weinberg-Tomozawa term,
given in chiral Lagrangian by
\cite{BROWN96A}
\begin{eqnarray}\label{vector}
V_{K^+} = {3\over 8f_\pi^2}\rho,
\end{eqnarray}
where $f_\pi$ is the pion decay constant and $\rho$ is the nuclear density.

The Weinberg-Tomozawa term is of leading order in the chiral expansion. 
In the next order come the Kaplan-Nelson term \cite{KAPLAN86} and the 
range term. The Kaplan-Nelson term can be expressed in terms of a 
scalar mean field acting on the kaon, either $K^+$ or $K^-$,
\begin{eqnarray}\label{scalar}
S_K =- {\Sigma _{KN} \over 2 m_K f_\pi^2} \rho _S.
\end{eqnarray}
In the above, $\rho _S$ is the nuclear scalar density and has a value of
about $\rho_S(\rho_0)\approx 0.93 \rho_0$ at normal nuclear matter density.
The $KN$ sigma term is denoted by $\Sigma_{KN}$, and its value depends
on the nucleon strangeness content and strange quark mass.

In studies of kaon production and flow in heavy-ion collisions
at SIS energies, two of the present authors and collaborators 
have used Eqs. (\ref{vector}) and (\ref{scalar}) for the kaon vector and
scalar potentials in nuclear matter \cite{FANG94,LI95A,LI95B,LI95C},
i.e., the kaon mean-field potential is approximately given by
\begin{eqnarray}\label{opti}
U_{K^+} \approx {3\over 8f_\pi^2} \rho - {\Sigma _{KN}\over 2m_K f_\pi ^2}
\rho _S .
\end{eqnarray}
Using $\Sigma _{KN} = 350$ MeV and $f_\pi$=93 MeV, the kaon potential
used in Refs. \cite{FANG94,LI95A,LI95B,LI95C} at normal nuclear matter 
density $\rho _0$ is about 7 MeV.  This is substantially smaller than 
that implied by the impulse approximation using the KN scattering length.
 
However, the pion decay constant changes in nuclear matter if we assume 
that the Gell-Mann$-$Oakes$-$Renner relation holds in medium, which is 
supported by model studies \cite{BROWN96B}.  Then we have
\begin{eqnarray}\label{gmor}
{f_\pi^{*2}\over f_\pi^2} = {m_\pi^2\over m_\pi^{*2}}
{\langle {\bar q}q\rangle ^*\over \langle {\bar q}q\rangle },
\end{eqnarray}
where quantities with ''*" denote values in medium.

From the Feynman-Hellmann theorem, one has \cite{COHEN92,LI94A}
\begin{eqnarray}
{\langle {\bar q}q\rangle ^*\over \langle {\bar q}q\rangle }
\approx 1-{\Sigma _{\pi N} \over f_\pi^2 m_\pi^2} \rho ,
\end{eqnarray}
where $\Sigma _{\pi N} =45 $ MeV is the pion-nucleon sigma term. The pion
mass is known to change only slightly as the density increases, so putting 
into Eq. (\ref{gmor}) the empirical value determined from $\pi$-mesonic
atoms \cite{EW88}, $m_\pi^*(\rho _0)/m_\pi \approx 1.05 $, one obtains
\begin{eqnarray}\label{fpi}
{f_\pi ^{*2}(\rho _0) \over f_\pi^2}\approx 0.6.
\end{eqnarray}

Since the rho meson mass is connected with the pion decay constant 
$f_\pi$ through the KFSR relation
\begin{eqnarray}\label{kfsr}
m_\rho^2 = 2 g_V^2 f_\pi^2, 
\end{eqnarray}
where $g_V$ is the universal vector meson coupling, a decreasing pion 
decay constant in medium implies that the in-medium rho meson is also 
reduced as $g_V$ changes only at loop level.  At mean-field level, 
Brown and Rho \cite{BROWN91} has thus introduced the following 
scaling relation,
\begin{eqnarray}\label{br}
{m_\rho^*\over m_\rho} \approx {m_\omega^*\over m_\omega}
\approx {f_\pi^*\over f_\pi}.
\end{eqnarray}
The dropping rho meson mass in dense matter is consistent with
recent transport model analyses \cite{LI95D,LI96A} of CERES \cite{CERES95}
and HELIOS-3 \cite{HELIOS95} dilepton data from CERN heavy-ion
experiments, which show enhanced production of dileptons in the mass
region between 300 and 500 MeV.

Including the scaling in $f_\pi$, kaon vector and scalar potentials
become
\begin{eqnarray}\label{vector1}
V_{K^+} = {3\over 8f_\pi^{*2}}\rho,
\end{eqnarray}
\begin{eqnarray}\label{scalar1}
S_K =- {\Sigma _{KN} \over 2 m_K f_\pi^{*2}} \rho _S.
\end{eqnarray}
For the $KN$ sigma term $\Sigma_{KN}$, Brown and Rho have shown in the
appendix of Ref. \cite{BROWN96B} that it has a value
\begin{eqnarray}\label{sigma}
\Sigma _{KN} \approx 450 \pm 30 ~{\rm MeV},
\end{eqnarray}
if the nucleon strangeness content is taken from recent lattice QCD
calculations \cite{LIU,FUKU}.  This value is somewhat larger than what 
has been used in Refs. \cite{FANG94,LI95A,LI95B,LI95C}.

Using $f_\pi^{*2}/f_\pi^2\approx 0.6$ and $\Sigma_{KN}\approx 450$ MeV,
then at normal nuclear matter density one has $V_{K^+}\approx 88$ MeV
and $S_K\approx 100$ MeV. In the Walecka model \cite{QHD}, the nucleon
mean-field potentials are $V_N\approx 270$ MeV and $V_S\approx
330$ MeV, one therefore has the approximate relation
\cite{BROWN96A,BROWN96B}
\begin{eqnarray}
V_{K^+} \approx {1\over 3} V_N, ~~~~~{\rm and}~~~~~
S_K \approx {1\over 3} S_N.
\end{eqnarray}
Thus, although
the Kaplan-Nelson term seems to be connected through $\Sigma _{KN}$
with the explicit breaking of chiral symmetry, an alternative
interpretation is that it gives a scalar field on the light quark
in the kaon.  We thus envisage that the constituent mass of
up or down quark in the kaon drops substantially in medium.
 
Starting from chiral Lagrangian with universal vector coupling,
Brown and Rho \cite{BROWN96A} have thus shown that the factor
$(0.6)^{-1}$ in Eq. (\ref{fpi}) just brings the scalar and vector 
mean fields in chiral Lagrangian to that of Walecka theory at nuclear
matter density. In a sense, they have derived, working through the
density dependence of $f_\pi^*$, the Walecka couplings at $\rho =\rho _0$,
beginning from the zero-density couplings in the chiral Lagrangian.
 
We assume that, as a working hypothesis, the vector and scalar mean fields
extrapolate linearly to densities higher than $\rho_0$, i.e., we
use $f_\pi^*(\rho _0)$ for higher densities. This is similar to
the Walecka model which is applied to higher densities with coupling 
constants determined at $\rho_0$. Of course, there should be some 
tendencies for the masses to continue to decrease, and the mean fields 
go as the inverse squared mass. However, this will be counteracted 
by higher order corrections, which tend to cut down the mean 
fields \cite{pand}.
 
In the same order of chiral perturbation theory as the Kaplan-Nelson
term is the `range term'. Much was made \cite{DEE92} of the fact that
this range term cancels off all of the attraction corresponding to
the Kaplan-Nelson-like term for the pion in pion-nucleon scattering.
This range term will also be important for a kaon. Detailed calculations
up through and including one loop order in chiral perturbation
theory \cite{LEE95} show that the range term can be taken into
account by multiplying the Kaplan-Nelson term by a factor
$1-0.37 \omega _K^2/m_K^2$, i.e.
\begin{eqnarray}\label{range}
(S_K)_{total} \approx -{\big[1-0.37 (\omega _K/m_K)^2\big]
\Sigma _{KN}\over 2m_K f^{*2}_\pi} \rho _S \approx - {\Sigma _{KN}
\over 2m_K f_\pi ^2 } \rho _S ,
\end{eqnarray}
The second expression arrives from the fact that the range term and 
the scaling in $f_\pi$ approximately cancel each other for the kaon.
 
The overall kaon potential including the scaling in $f_\pi$ and 
the range term is then 
\begin{eqnarray}\label{opti1}
U_{K^+} \approx {3\over 8f_\pi^{*2}} \rho - {\Sigma _{KN}\over 2m_K f_\pi ^2}
\rho _S .
\end{eqnarray}
With again $\Sigma _{KN}$=450 MeV, the kaon potential at $\rho _0$ from
Eq. (\ref{opti1}) is about 28 MeV, which agrees very well with
the empirical value of 30 MeV. Actually the two agree with each other
up to 2$\rho _0$ as shown in Fig. 1. We also show in Fig. 1 the kaon
potential in the case of vanishing scalar attraction.
 
\section{kaon production and flow at SIS}

Experimental measurement of kaon production in Au+Au collisions
at 1 GeV/nucleon has been carried out by the KaoS collaboration
at SIS \cite{KAOS}.  Both Fang {\it et al.} \cite{FANG94} and
Maruyama {\it et al.} \cite{MARU94} have analyzed the KaoS data
using relativistic transport models. In Ref. \cite{FANG94} a good fit to
the data was obtained by using the kaon potential given by Eq.
(\ref{opti}) without the range term and medium modification
of $f_\pi$. As mentioned above, the kaon potential used in
Ref. \cite{FANG94} is less repulsive than both the one from the 
impulse approximation and the one given by Eq. (\ref{opti1}) as 
suggested by Brown and Rho based on effective chiral Lagrangian with 
scaling pion decay constant.
 
Maruyama {\it et al.} \cite{MARU94} have independently obtained
a good fit to the kaon yield by putting the same Walecka-type
scalar mean field on the lambda hyperon as on the nucleon and delta 
resonance but none on kaon. However, with our rule of nonstrange
quark counting, 2/3 of the scalar field should be applied to the 
lambda hyperon and 1/3 to kaon as was done by Fang {\it et al.} 
\cite{FANG94}. The threshold is thus quite similar in both cases.
According to Randrup and Ko \cite{KO80}, kaon production at a given
input energy is chiefly determined by the maximum kaon momentum $p_{max}$,
which in turn is determined by the threshold. It is therefore
quite natural that both Fang {\it et al.} and Maruyama {\it et al.}
could fit equally well the measured kaon yield.
 
In this work we recalculate the kaon spectra in Au+Au collisions 
at 1 GeV/nucleon with the kaon potential given by Eq. (\ref{opti1}) 
and including also the contributions from pion-nucleon interactions.
The isospin-averaged cross sections for $\pi N\rightarrow 
\Lambda K$ and $\pi N\rightarrow \Sigma K$ are taken from 
Ref. \cite{CUG84}. The results are shown in Fig. 2 where the 
solid and dashed curves correspond to the results obtained 
with ($\Sigma _{KN}$ =450 MeV) and without ($\Sigma _{KN}$=0 MeV) 
kaon scalar potential. Although the larger repulsive kaon potential 
used in the present calculation reduces kaon yield from
baryon-baryon interactions compared to that of Ref. \cite{FANG94}, it is,
however, compensated by contributions from the pion-nucleon interactions.
The present results with a more repulsive kaon potential than the one 
used in Ref. \cite{FANG94} thus still agree very well with the 
experimental data. On the other hand, if we neglect the kaon scalar 
potential, the theoretical results are reduced by about a factor of 2-5.
 
Another piece of useful information about $K^+$ properties in dense
matter has recently been provided by the FOPI collaboration \cite{FOPI}
from measuring the kaon flow in Ni+Ni collisions at 1.93 GeV/nucleon.
Li {\it et al} \cite{LI95B,LI95C} have shown that the FOPI data can be well
fitted using the vector and scalar mean-field potentials given by
Eqs. (\ref{opti}) with $\Sigma_{KN}=350$ MeV and $f_\pi=93$ MeV.
 
We have redone the calculation using Eq. (\ref{opti1}) with
$\Sigma_{KN}=450$ MeV and scaling pion decay constant. The results are 
shown in Fig. 3 together with the preliminary FOPI data \cite{FOPI}. 
Three scenarios for the kaon potential in nuclear matter have been 
considered, i.e., without potential (dotted curve), with vector potential 
only (dashed curve), and with both scalar and vector potentials 
(solid curve). It is seen that with both larger vector and scalar 
potentials than those in Refs. \cite{LI95B,LI95C}, the theoretical 
results are still consistent with the FOPI data. On the other hand, 
if the kaon scalar potential is neglected, then there appears a clear
antiflow of kaons with respect to nucleons, which is seen to contradict 
the data.  Since the magnitude of kaon vector potential seems secure,
the lack of kaon antiflow in the experimental data thus indicates that
there is an appreciable scalar attraction for kaon in order to
cancel partly its strong repulsive vector potential.
 
\section{kaonic atom}

For $K^-$ in nuclear medium, because of G-parity, the vector potential 
is attractive,
\begin{equation}
V_{K^-}=-\frac{3}{8f_\pi^{*2}}\rho.
\end{equation}
The $K^-$ scalar potential is also given by Eq. (\ref{range}) but
with a less important range term than for kaon. Taking, roughly, 
$\omega _{K^-}$ = 295 MeV corresponding to the 200 MeV binding found 
by Friedman, Gal, and Batty at $\rho\approx \rho _0$ for a $K^-$ meson 
around a $^{56}$Ni atom \cite{FGB93}, we find $0.37 (\omega _{K^-}/m_K)^2 
=0.13 $, so the correction for the range term is only about 13\%. 
Taken literally, this would mean that at $\rho \approx \rho_0$,
\begin{eqnarray}
S_{K^-} (\rho _0) \approx -94~{\rm MeV} {\rho _S\over \rho _0}.
\end{eqnarray}
Adding vector and scalar interactions gives
\begin{eqnarray}
U_{K^-} (\rho _0) = S_{K^-} (\rho _0) + V_{K^-} (\rho _0) \approx
-94~{\rm MeV} {\rho_S\over \rho_0}
- 88~{\rm MeV} \approx - 175~{\rm MeV}.
\end{eqnarray}
The magnitude of attraction is thus consistent with the $-200\pm 20 $ 
MeV found by Friedman, Gal and Batty \cite{FGB93}.

Consequently, although the attraction in the scalar mean field needed 
to produce the observed number of kaons and flow in heavy-ion collisions
looks small (about -65 MeV$\rho _S/\rho _0$), compared with that found 
in kaonic atoms (about -94 MeV $\rho _S/\rho_0$), once the correction 
is made including the range term, the scalar attraction needed in the 
two cases is very similar and is consistent with the one given by 
Kaplan and Nelson but corrected by a scaling of the pion decay constant.

We note that once the $K^-$ mean-field potential is properly included, 
then the effect of $\Lambda(1405)$, which can be considered as a bound 
$\bar KN$ state, is suppressed in medium as a result of large energy 
denominator and Pauli blocking effects \cite{LEE}.

\section{conclusions}

Our conclusion is that basically the nonstrange quark in the
kaon feels about 1/3 of the scalar and vector mean field potentials
that a nucleon does in the Walecka model. However, corrections for 
the energy-dependent range term must be made, which is significant 
for $K^+$, but relatively unimportant for $K^-$. For the vector mean 
field our argument that $V_K \approx {1\over 3} V_N$, seems quite 
secure, depending only on the fact that the coupling goes as the 
nonstrange baryon numbers. For the scalar mean field, both the 
enhancement of subthreshold kaon yield and the lack of kaon 
antiflow in heavy-ion collisions at the SIS energies are found 
to be consistent with the scenario that it is somewhat larger in 
magnitude than the vector mean field before being cut down by 
the range term.
 
In this way one can also understand the very large attraction,
$V_{K^-} = -200\pm 20 ~{\rm MeV}$, found by Friedman, Gal and Batty 
\cite{FGB93} for a $K^-$ in $^{56}$Ni. Furthermore, the sum of absolute 
values of scalar and vector mean fields, about 600 MeV, that
enters into the spin-orbit term in nuclei in the Walecka theory,
turns out to be about three times the $K^-$ potential. We thus believe 
that our arguments and interpretation of the KaoS and FOPI data give 
important information for the attractive scalar interaction acting on 
a kaon in dense matter, which is needed in the calculation of kaon 
condensation in stars \cite{LEE95,BROWN96B}.
 
\vskip 1cm
 
We are grateful to D. Best, E. Grosse, and J. Ritman for useful communications.
GEB is supported by the Department of Energy under Grant No. DE-FG02-88ER40388,
and CMK and GQL are supported by the National Science Foundation
Grant No. PHY-9509266.

\newpage
 
\centerline{\bf Figure Captions}
 
Fig. 1:  ~~Kaon potential as a function of density from Eq. (\ref{impulse})
and Eq. (\ref{opti1}) with $\Sigma _{KN}$=450 and 0 MeV, respectively.
 
Fig. 2:  ~~Kaon momentum spectra from Au+Au collisions at 1 GeV/nucleon.
The experimental data are from Ref. \cite{KAOS}.
 
Fig. 3:  ~~Kaon transverse momentum as a function of center-of-mass 
rapidity for Ni+Ni collisions at 1.93 GeV/nucleon and b$\le$ 4 fm. The
preliminary experimental data are from Ref. \cite{FOPI}.
 
\end{document}